\documentclass[aps,prl,preprint,superscriptaddress]{revtex4-2}

\usepackage{mhchem}
\usepackage{xcolor}
\usepackage{graphicx}
\usepackage{multirow}
\usepackage{hyperref}
\hypersetup{
    colorlinks=true,
    linkcolor=blue,
    urlcolor=blue,
    citecolor=blue,
    }
\usepackage[outdir=./]{epstopdf}

\begin{document}


\title{High-Resolution Spectroscopy of the X-A Transition of the Carbon Monoxide Dication \ce{CO^{2+}}}


\author{X.\,Huet}
\email[]{xavier.huet@ulb.be}
\author{A. Aerts}
\author{N. Vaeck}
\affiliation{Spectroscopy, Quantum Chemistry and Atmospheric Remote Sensing (SQUARES), Universit\'e libre de Bruxelles, 50 Avenue F. Roosevelt, C.P. 160/09, B-1050 Brussels, Belgium}

\author{M.\,G\'en\'evriez}
\email[]{matthieu.genevriez@uclouvain.be}
\author{X.Urbain}
\email[]{xavier.urbain@uclouvain.be}
\affiliation{Institute of Condensed Matter and Nanosciences, Universit\'e catholique de Louvain, BE-1348 Louvain-la-Neuve, Belgium}


\date{\today}

\begin{abstract}
We report rovibronic spectra of the A $^3\Sigma^+$($v'=0-2$) - X $^3\Pi_\Omega(v=0)$ rovibronic transitions ($|\Omega|=0, 1$ and 2) of the CO$^{2+}$ doubly-charged molecular ion. Spectra were recorded at high resolution ($\sim 5$~cm$^{-1}$) in a fast beam of CO$^{2+}$ molecules by detecting the Coulomb explosion of the molecules upon excitation to the A state. Measurements were guided by \textit{ab initio} calculations which then assisted the assignment of the observed spectral features. Our results resolve the spin-orbit splittings of the ground vibronic state X $^3\Pi_\Omega(v=0)$, but not the rotational structure of the bands due to spectral congestion, and provide spectroscopic information on CO$^{2+}$ with unprecedented resolution. In doing so they expand our knowledge of this benchmark doubly charged molecular ion and expand the short list of doubly charged molecules studied at high resolution. 
\end{abstract}


\maketitle

\section{Introduction}\label{Introduction}

\begin{figure}
 \includegraphics{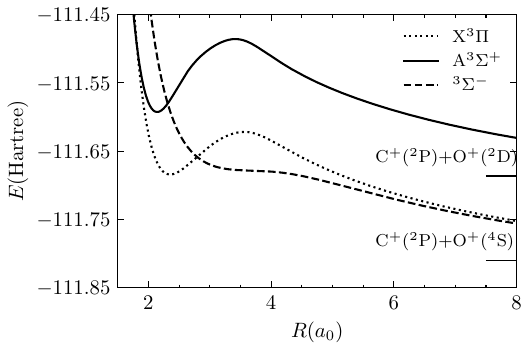}%
 \caption{\label{pecs4}Potential electronic curves of \ce{CO^{2+}} that illustrate the peculiar topology commonly encountered for low-lying electronic states in diatomic dications. Both X$^3\Pi$ and $^3\Sigma^+$ electronic states dissociate into C$^+$($^2$P)+O$^+$($^4$S) fragments and the $^3\Sigma^-$ electronic state dissociates into C$^+$($^2$P)+O$^+$($^2$D) fragments.}
 \end{figure}

Diatomic dications \ce{AB^{2+}} are intriguing and elusive chemical compounds. They
exhibit a broad range of structure, from thermodynamically stable to metastable
and unstable, depending on the single- and double-ionization energies of their
atomic constituents A and B~\cite{schroder99,sabzyan2014diatomic,mathur93}. Many diatomic dications of atmospherical and
astrophysical interest~\cite{thissen2011doubly} are thermodynamically unstable but possess low-lying electronic states whose
potential-energy curves (PECs) exhibit a local well
embedded in the repulsive \ce{A+}+\ce{B^+} Coulomb barrier, as illustrated in Fig.~\ref{pecs4} for the case of CO$^{2+}$. The well supports metastable vibrational levels lying several
electronvolts above the
\ce{A+}+\ce{B^+} dissociation limit and exhibiting lifetimes exceeding the
seconds range~\cite{sabzyan2014diatomic,mathur95}. The lifetime of these states is
affected by the tunneling through the outer potential barrier of the well and
nonadiabatic couplings to other electronic states~\cite{andersen1993very,mrugala2008computational}. The combination of their
large internal energy, high reactivity and long lifetime means that diatomic
dications could play a significant role in atmospheric
physics~\cite{thissen2011doubly,falcinelli16,alagia13,gu20,cheng25,lilensten13}.

Among molecular dications CO$^{2+}$ plays a special, benchmark role. It is the
first long-lived dication ($\tau > 1$~$\mu$s) ever detected
experimentally~\cite{conrad1930appearance}. The lifetimes of its metastable
levels have since proved to be remarkably sensitive benchmarks to test quantum
chemical predictions and, in particular electronic potentials, nonadiabtic
spin-orbit couplings, and numerical methods. The peculiar topology of the
ground-state PEC was identified early on~\cite{wetmore1984theoretical} but the
proposal of tunneling as the main dissociation pathway was later called into
question~\cite{larsson1989theoretical} and finally
disproved~\cite{andersen1993very}. The latter work showed that spin-orbit
couplings open dissociation pathways for three of the lowest lying states, and
recent analyses~\cite{vsedivcova2006computed,
vsedivcova2007radiative,mrugala2008computational} showed extensively and with
exquisite detail how spin-orbit couplings play, in fact, a predominant role in the
predissociation of \ce{CO^{2+}}. Experimentally, the lifetime of metastable
CO$^{2+}$ has been investigated at storage rings, yielding a value of $> 3.8$~s
for the ground vibronic level, and recent modelling of the Martian atmosphere
suggests it could significantly exceed this lower bound~\cite{cheng25}. Other levels with long lifetimes
($\geq 0.2$~ms) have also been observed~\cite{mathur95} although no definitive
assignment could be made.

As is the case for most diatomic dications, the rovibronic structure of
CO$^{2+}$ is known only at relatively low resolution ($> 10$~meV). Spectroscopic information was obtained in experiments where neutral CO are doubly ionized and the fragments,
either electrons~\cite{dawber1994threshold,eland03}, ions~\cite{curtis1984ion,lundqvist1995novel,bouhnik2001measurements}, or combinations of the two~\cite{
penent1998new}, are detected in coincidence. Starting from these
low-resolution data several attempts were made to record the X-A
transition at higher resolution with laser light, but remained
unsuccessfull~\cite{cox2003high,eland03}. A tentative observation of two rovibronic lines of the A($v'=0$)-X($v=0$) band has been reported~\cite{cossart99} but has since been
challenged by theory~\cite{mrugala2008computational} and awaits further
experimental verification.

The lack of rotationally resolved data for CO$^{2+}$ is certainly not an
isolated situation for diatomic dications and, to date, only
\ce{NO^{2+}} \cite{cossart1987optical}, \ce{N2^{2+}}
\cite{carroll1961identification, cosby1983photofragment, cossart1985optical,
mullin1992triplet, martin1994photofragment}, \ce{DCl^{2+}} \cite{cox2003high}
and \ce{MgAr^{2+} \cite{wehrli2021spectroscopic}} have been studied
with rotational resolution. The scarcity of spectral data, in particular the lack of precisely known emission lines, is an important bottleneck for studying dications in atmospheric and astrophysical environments and for benchmarking quantum-chemical calculations of the structure and subtle decay dynamics of dications.

In an effort to expand our knowledge of molecular dications, the present work
reports an experimental investigation, supported by \textit{ab initio}
calculations, of the rovibronic spectrum of \ce{CO^{2+}} at high resolution.
We recorded the spectra of the A($v'=0-2$)-X($v=0$) vibrational bands at a
resolution of $\sim 5$~cm$^{-1}$, yielding band origins and spin-orbit
splittings of the ground electronic state with an accuracy improved 25-fold compared to existing data. As described in Sec.~\ref{sec:experiment},
the spectra were recorded by exciting ground-state CO$^{2+}$ molecules with
laser light and utilizing their fast predissociation in the excited A($v'$)
state to detect the C$^+$ and O$^+$ fragments in coincidence. The use of laser excitation yields a spectral resolution that is significantly improved compared to previous works on the same dication~\cite{dawber1994threshold,eland03, curtis1984ion,lundqvist1995novel,bouhnik2001measurements, 
penent1998new, cox2003high}, and the use of a coincidence detection technique enables us to extract the relevant, weak predissociation signal from a large background of other nonresonant dissociations (see  Ref. \cite{cox2003high} for a discussion on other species). Experimental
searches were assisted by high level quantum chemistry calculations targeted at
prediciting accurate wavenumbers for the X-A transition and a comprehensive
picture of spin-orbit couplings and predissociation lifetimes
(Sec.~\ref{sec:abinitio}). Experimental spectra and theoretical results are presented and analyzed in Sec.~\ref{sec:results}.  


\section{Experiment}\label{sec:experiment}

\begin{figure}[h!]
  \includegraphics[width=0.8\linewidth]{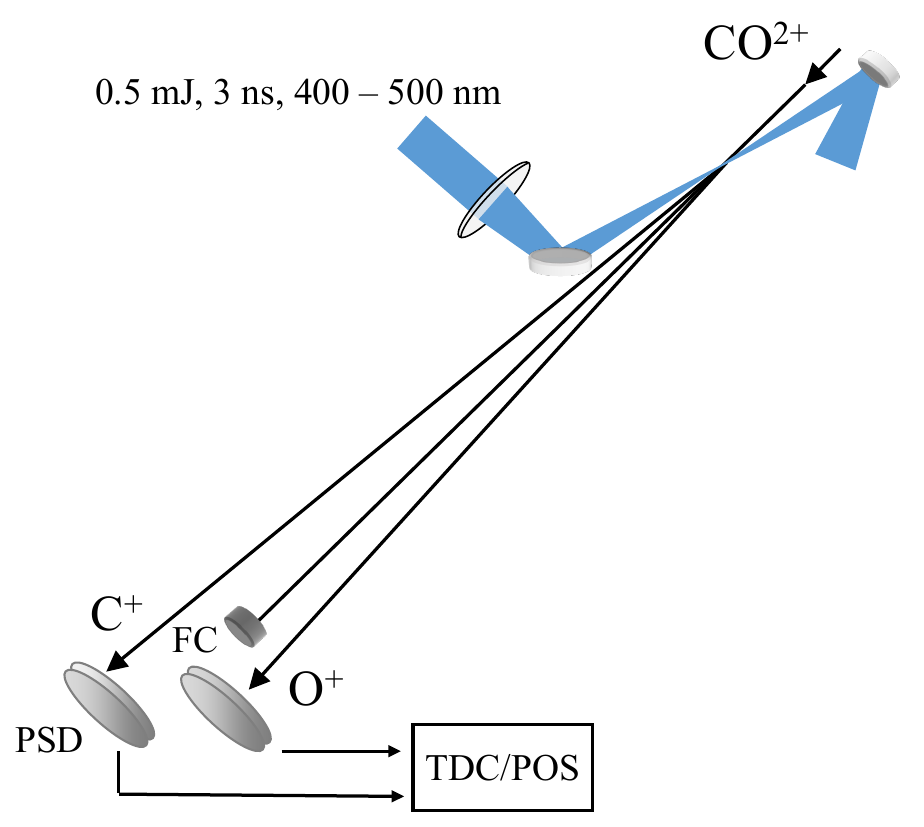}
  \caption{Scheme of the spectroscopy and detection regions. Fast CO$^{2+}$ ions are photodissociated by a near-collinear laser beam and the C$^+ $ and O$^+$ fragments are detected in coincidence by a pair of position sensitive detectors (PSD) allowing simultaneous determination of the impact positions (POS) and arrival times using a time-to-digital converter (TDC)}. Primary ions ares collected by a Faraday cup (FC).
  \label{experiment}
\end{figure}

The experimental setup derives from that used in previous photodissociation studies on the H$_3^+$ ion \cite{Urbain2019}. The apparatus comprises an ion source, an acceleration column and a mass-to-charge magnetic selector. The source is a conventional discharge source (duoplasmatron) operated with CO$_2$ gas at the lowest pressure achievable to maintain the arc, \textit{i.e.} $10^{-2}$~mbar. Under such conditions a typical current of 500~pA is formed into a 5~keV beam collimated by a pair of apertures (2~mm and 1~mm diameter, respectively). The beam is translated parallel to its original propagation axis just before the interaction with the laser beam to limit the contribution of dissociations happening along the flight path. The laser beam is produced by an optical parametric oscillator (OPO) pumped by the third harmonic of a pulsed Nd:YAG laser operating at 30~Hz repetition rate and delivering pulses of 3~ns duration ({\it Ekspla} NT342A). In order to avoid saturation of the photodissociation process, we run the laser beam nearly parallel to the ion beam (Fig. \ref{experiment}). The angle between ions and laser beam ($10\pm 5$ degree) results in a 19.2(4) cm$^{-1}$ Doppler shift. The photodissociation products C$^+$ and O$^+$ are detected 1.7~m downstream by a pair of position-sensitive detectors operating in coincidence, which consist of a triple-microchannel plate stack backed with a resistive anode for position encoding ({\it Quantar} RAE 3300 Series). The parent CO$^{2+}$ are collected by a Faraday cup.

From the known origin of the photodissociation products, their time of flight and positions of arrival, we compute the total kinetic energy release (KER). We make use of  total momentum and energy conservation:
\begin{eqnarray}\label{momentum}
  M\vec{V}&=&m_1\vec{V}_1+m_2\vec{V}_2,\\
  M^2 V^2&=&m_1^2 V_1^2+m_2^2 V_2^2+ 2 m_1 m_2 \vec{V}_1\cdot\vec{V}_2
\end{eqnarray}
where $m_1$, $m_2$ are the fragment masses, $\vec{V_1}$, $\vec{V_2}$ their velocities in the laboratory frame, and $M$, $\vec{V}$ the parent ion mass and velocity. Indices 1/2 refer to either detector, and correspond to C$^+$/O$^+$ or O$^+$/C$^+$, their identification resting on the reconstructed point of impact in the detector plane being close to the beam propagation axis. Defining the displacement of either particle, $\vec{d}_i=\vec{V}_i t_i$, we may recast the non-linear Eq. \ref{momentum} to be solved for $t_1$, the time-of-flight of the particle hitting detector 1:
\begin{align}\label{timeofflight}
  M^2 V^2&=
  m_1^2 \frac{d_1^2}{t_1^2}+m_2^2 \frac{d_2^2}{(t_1+\tau)^2}\nonumber\\
  &+ 2 m_1 m_2 \frac{d_1 d_2}{t_1(t_1+\tau)} \cos(\vec{d}_1,\vec{d}_2),
\end{align}
making use of the fact that $t_2=t_1 +\tau$, with $\tau$ the recorded time-of-flight difference between detectors. We finally obtain $E_K$, the value of the KER:
\begin{equation}
  E_K=\frac{1}{2}(m_1 v_1^2+ m_2 v_2^2)
\end{equation} 
where $\vec{v}_i=\vec{V}_i-\vec{V}$ are the velocities in the center-of-mass frame.

\begin{figure}[h!]
  \includegraphics[width=\linewidth]{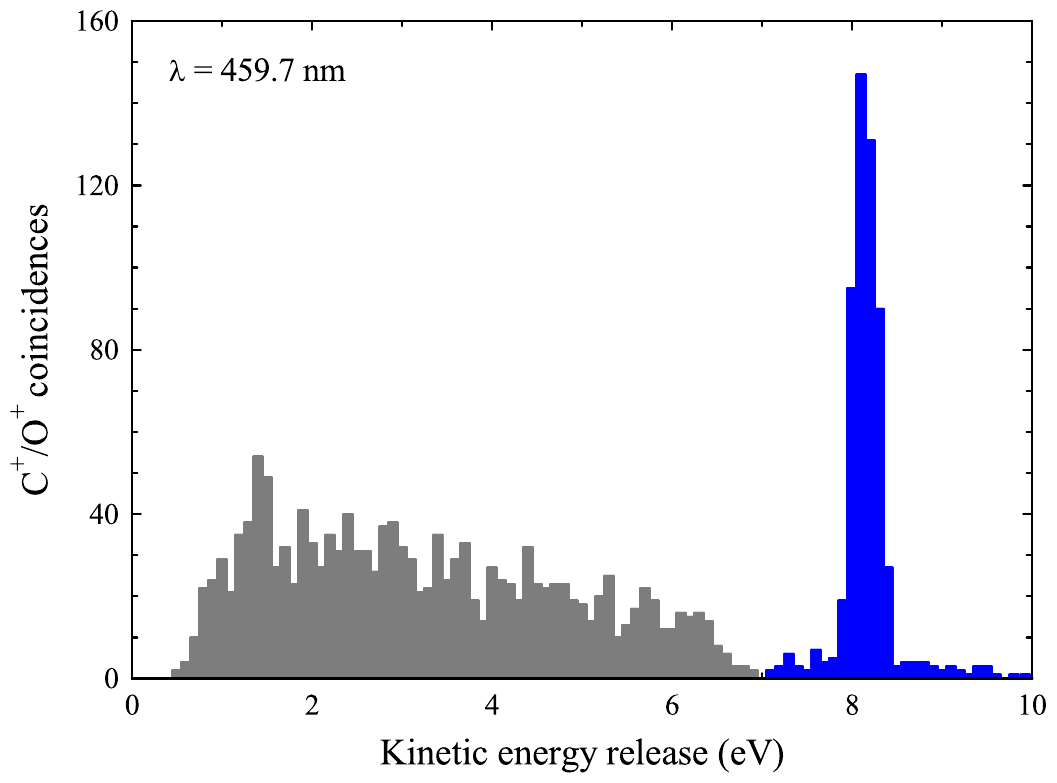}
  \caption{Distribution of C$^+$-O$^+$ coincidences as a function of total kinetic energy release. The sharp peak near 8 eV corresponds to the photodissociation of CO$^{2+}$ molecules in the X state, while the broad distribution corresponds to spontaneous dissociation events along the beam flight toward the detectors. The blue area shows the event-selection window, the gray area the background events used for normalization.}
  \label{KER}
\end{figure}

 The laser-induced KER amounts to the energy difference between the CO$^{2+}$ ground state and the ${\rm C}^+ + {\rm O}^+$ asymptote, augmented by the photon energy. The width of the peak (blue area in Fig. \ref{KER}) is $\Delta E_\text{K} = 0.3$~eV, and is mainly due to the extent of the laser beam overlap, $\sim 3$ cm, and remains constant for all the data points shown below. The characteristic KER, together with the time-of-flight of the fragments referenced to the laser pulse, allows us to distinguish the laser induced events from the spontaneous dissociation of CO$^{2+}$ occurring everywhere along the beam trajectory. The KER of such events is inferior to that of photodissociation events by the photon energy. However, an additional difficulty arises from the exact location of the spontaneous dissociation being unknown, causing the computed kinetic energy to over- or under-estimate the actual kinetic energy release depending on whether the dissociation took place upstream or downstream of the laser crossing point. This is the origin of the broad distribution observed to extend between 0.5 and 6.5~eV (gray area in Fig. \ref{KER}). 

The upstream contribution is limited due to the presence of the chicane created by the lateral displacement of the beam right before the interaction point, allowing for total separation of the photodissociation and spontaneous dissociation signals. The actual coincidence rate of spontaneous events exceeds by far that of laser-triggered events. To alleviate that problem, the beam is chopped into 200~ns long bunches synchronized with the laser pulses by means of the chicane deflectors, and the beam intensity is adjusted to keep the coincidence rate well below one dissociation per laser pulse. An additional time selection encompassing the photodissociation signal is operated before computing the histogram presented in Fig. \ref{KER}. The photodissociation signal is normalized to the spontaneous dissociation background, both being proportional to the ion beam intensity. This normalized signal is further divided by the laser pulse energy as a proxy for the photodissociation cross section.

\section{\textit{Ab initio} calculations}\label{sec:abinitio}
An independent theoretical \textit{ab initio} investigation is carried out alongside the experimental study to determine the optimal wavelength for exciting the \ce{CO^{2+}} dication and inducing its predissociation. To this end, the rovibronic levels of the molecule are computed, along with their lifetimes, to construct a realistic theoretical absorption cross-section spectrum for the transition of interest.

\subsection{Electronic states and SO coupling}
The potential energy curves (PECs) $E^{el}(R)$ of the lowest-lying electronic states of \ce{CO^{2+}} have been computed by solving the electronic Schr\"odinger equation,
\begin{equation}\label{el-sch-eq}
    \hat{H}^{el}(R)\Psi(R)\:=\: E^{el}(R)\Psi(R),
\end{equation}
with the \texttt{MOLPRO} package \cite{MOLPRO}, using CASSCF \cite{WK85} and MRCI \cite{WK88} methods. In Eq. \ref{el-sch-eq}, $\hat{H}^{el}(R)$ is the electronic Hamiltonian,  $\Psi(R)$ the electronic wavefunction and $R$ the internuclear distance. The electrons are distributed in the molecular orbitals as follows: the 2 lowest $\sigma$ orbitals are systematically kept doubly occupied and the remaining of the electrons are distributed among the 4 $\sigma$ and 4 $\pi$ above orbitals. The same active space was used for both CASSCF and MRCI calculations. In order to adequately describe the core orbitals, Dunning's augmented correlation consistent core-valence orbital basis set aug-CC-pCV6Z (ACV6Z) \cite{WILSON1996339} has been employed for both carbon and oxygen. 

\begin{figure}
 \includegraphics{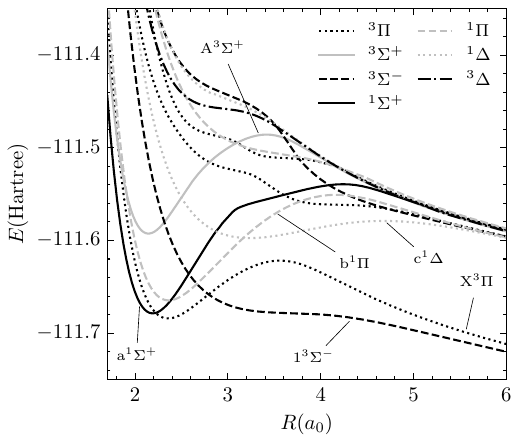}
 \caption{\label{PECs}Potential energy curves of \ce{CO^{2+}} computed at the CASSCF-MRCI/aug-cc-pCV6Z level of theory. Electronic states important to the discussion are pointed out.}
 \end{figure}

The PECs of twelve electronic states have been calculated and are presented in Fig. \ref{PECs}. Each curve is obtained from the electronic energies calculated at 98 internuclear distances $R$, ranging from 1.7 to 15 $a_0$, and then fitted with cubic B-spline functions. Each electronic state is identified by its spin-multiplicity and its symmetry in the C$_{\infty v}$ point group. Spectroscopic constants of selected electronic state computed with the DIATOMIC routine of the \texttt{MOLPRO} package \cite{MOLPRO} are presented in Table \ref{spec_cst}. They allow comparison of the topology between the present PECs and those reported in previous studies. Our results are consistent with those provided for comparison, with particularly good agreement with those from Eland \textit{et al.} \cite{eland2004photo}. The root-mean-square errors of $R_{\text{e}}$, $B_e$, $\omega_e$, and $\omega_e x_e$, which quantify the deviation between our computed values and those predicted in previous studies, are $1.5 \times 10^{-3}$ $a_0$, $1.6 \times 10^{-3}$ cm$^{-1}$, 14.4 cm$^{-1}$, and 1.38 cm$^{-1}$ for \cite{eland2004photo} and $1.1 \times 10^{-2}$ $a_0$, $1.5 \times 10^{-2}$ cm$^{-1}$, 24.9 cm$^{-1}$ and 1.83 cm$^{-1}$ for \cite{vsedivcova2006computed}. Discrepancies between our results and those previously reported can be attributed to the choice of basis set and active space. Eland \textit{et al.} \cite{eland2004photo} used Dunning’s V5Z basis set, whereas Šedivcová \textit{et al.} \cite{vsedivcova2006computed} used the V6Z one. Both of those studies considered an active space containing the two lowest-energy molecular orbitals, \textit{i.e.} they were not kept closed, but considered two fewer $\pi$ active orbitals as compared to us. It seems surprising that our results are closer to those obtained with a smaller basis set. The explanation we propose is that Dunning’s VXZ and AVXZ basis sets are less suited compared to ACVXZ for studying this system with such an active space, as there is a significant imbalance between the description of core and valence orbitals. This imbalance becomes even more pronounced as the cardinal number X increases. With the active space used in \cite{eland2004photo} and \cite{vsedivcova2006computed}, the results of the multi-configurational calculations are particularly affected by this imbalance, as no orbital is closed. 
\\

 \begin{table}
\caption{\label{spec_cst}Spectroscopic constants of PECs of the lowest-lying electronic states of \ce{CO^{2+}}. Values are in cm$^{-1}$ unless otherwise stated. $R_{\text{e}}$, $B_e$, $\omega_e$, and $\omega_e x_e$ are the distance of equilibrium, the rotational constant, the vibrational constant and the anharmonicity constant, respectively.}
\begin{ruledtabular}
\begin{tabular}{cccccc}
 & Ref. & $R_{\text{e}}$ ($a_0$) & $B_e$ & $\omega_e$  & $\omega_e x_e$  \\
\hline
X$^3\Pi$      & This study & 2.356 & 1.581 & 1442.41 & 24.68 \\
& \cite{vsedivcova2006computed} & 2.342 & 1.599 & 1480.3 & 23.2 \\
& \cite{eland2004photo} & 2.358 & 1.579 & 1435.2 & 24.6 \\
\hline
A$^3\Sigma^+$ & This study & 2.144 & 1.910 & 2050.92 & 23.61 \\
& \cite{vsedivcova2006computed} & 2.137 & 1.919 & 2068.6 & 20.3 \\
& \cite{eland2004photo} & 2.144 & 1.909 & 2043.8 & 16.3 \\
\hline
a$^1\Sigma^+$ & This study & 2.191 & 1.831 & 1947.37 & 17.24 \\
& \cite{vsedivcova2006computed} & 2.182 & 1.844 & 1951.3 & 17.6 \\
& \cite{eland2004photo} & 2.190 & 1.829 & 1920.5 & 16.7 \\
\hline
b$^1\Pi$      & This study & 2.360 & 1.574 & 1490.55 & 17.52 \\
& \cite{vsedivcova2006computed} & 2.349 & 1.591 & 1517.5 & 17.3 \\ 
& \cite{eland2004photo} & 2.362 & 1.573 & 1487.7 & 17.2
\end{tabular}
\end{ruledtabular}
\end{table}

\begin{figure}
 \includegraphics{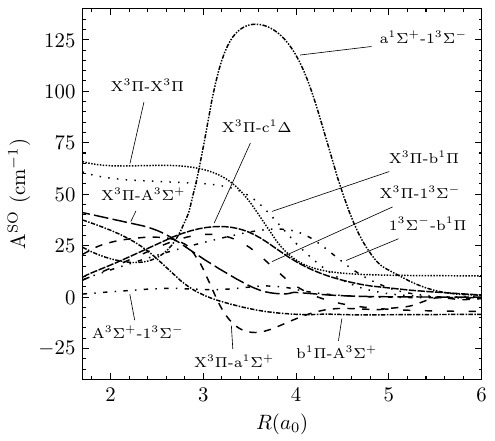}%
 \caption{\label{aso}Spin-orbit coupling between the lowest-lying electronic states of \ce{CO^{2+}}.}
 \end{figure}

Six electronic states are of particular interest in the remainder of the study and are pointed out in Fig. \ref{PECs}. Five of them exhibit PECs with unusual topology, as explained in the introduction: a potential barrier allows the presence of metastable rovibronic states within the well of the curve. 
\\\\
The spin-orbit coupling matrix elements (SOMEs) were computed for the electronic states X$^3\Pi$, a$^1\Sigma^+$, A$^3\Sigma^+$, b$^1\Pi$, c$^1\Delta$, and 1$^3\Sigma^-$ using the dedicated routine of the MRCI program \cite{berning2000spin} implemented in the \texttt{MOLPRO} package. The SOMEs are defined as \cite{lefebvre2004spectra}:  
\begin{equation}\label{aso_def}
  \text{A}_{ij}^{\text{SO}} \:=\: \langle \Psi^{el}_i | \hat{H}^{\text{SO}} | \Psi^{el}_j \rangle,
\end{equation}
where $\hat{H}^{\text{SO}}$ is the spin-orbit Hamiltonian defined by the Breit-Pauli (BP) operator and indices $i$ and $j$ run over the set of states listed above. Moreover, the AV5Z basis set was chosen for the computation of the SOMEs as the difference with those computed with the ACV6Z basis set were found to fall below the intrinsic accuracy of the theoretical method, which is 1 cm$^{\text{-1}}$\cite{berning2000spin}. The SOMEs were evaluated at the same points as the PECs, and each element fitted to cubic B-spline. The results are presented in Fig. \ref{aso}.\\

It is worth to note that some of the spin-orbit interactions are strictly zero because only electronic states with the same $\Omega = \Sigma + \Lambda$ value interact via homogeneous spin-orbit coupling. Here, $\Sigma$ is the projection of the spin angular momentum, and $\Lambda$ is the projection of the electronic orbital angular momentum, both onto the internuclear axis. \\

By accounting for this relativistic correction, the energy gap between the $\Omega$ components of a single electronic state $i$ is given by $\text{A}_{ii}^{\text{SO}} \Lambda \Sigma$. Among the six electronic states mentioned, only the $\Omega$ components of X$^3\Pi$ are expected to exhibit a lifting of degeneracy. This energy shift amounts to approximately $\pm$64 cm$^{-1}$ in the region of the potential wells. The vibrational structure of X$^3\Pi_{\Omega=0}$, X$^3\Pi_{|\Omega|=1}$, and X$^3\Pi_{|\Omega|=2}$ in the region is then expected to be very similar but shifted by the same energy gap. Note that the spin-orbit coupling element $\text{A}_{\text{X}^3\Pi\text{-X}^3\Pi}^{\text{SO}}$, is also presented in Fig. \ref{aso}.

\subsection{Vibrational levels}
The vibrational levels and their lifetime were computed in two successive steps. Each electronic state is first treated individually with the finite-element discrete-variable representation method combined with exterior complex scaling (FEM-DVR-ECS). The method has been described in detail in several works, in particular by Rescigno and McCurdy \cite{rescigno2000numerical} and by Génévriez \textit{et al.} \cite{genevriez2019experimental}. For a given adiabatic electronic state we solve the nuclear Schr\"odinger equation,
\begin{equation}
    \left[ \hat{T}(R) \:+\: V(R) \right]|\chi_j(R)\rangle = E_j |\chi_j(R)\rangle,
\end{equation}
where $\chi_j(R)$ are the vibrational wavefunctions with associated energies $E_j$, $V(R)$ the potential energy curves $E^{el}(R)$ of Fig.~\ref{PECs} corrected by the centrifugal barrier, $\hat{T}(R)$ the nuclear kinetic energy operator, 
\begin{eqnarray}
    \hat{T}(R)\:=\:\frac{-1}{2\mu}\frac{\text{d}^2}{\text{d}R^2},
\end{eqnarray}
and $\mu$ the reduced mass of the system.
\\\\
Off-diagonal SOMEs coupling different electronic states are taken into account in the second step by solving the complete Schr\"odinger equation
\begin{equation}\label{schrodinger_total}
    \left[\hat{H}^{el}(R) + \hat{T}(R) + \hat{H}^\text{SO}(R)\right]|\Psi^{el}_i\rangle |\chi^{(i)}_j(R)\rangle = \varepsilon |\Psi^{el}_i\rangle |\chi^{(i)}_j(R) \rangle,
\end{equation}
where index $i$ runs over the electronic states for which SOMEs have been computed, namely X$^3\Pi$, a$^1\Sigma^+$, A$^3\Sigma^+$, b$^1\Pi$, c$^1\Delta$, and 1$^3\Sigma^-$. This results in a final matrix of size 4792\,$\times$\,4792 to be diagonalized. Such a methodology~\cite{genevriez20a,wehrli21b} enables the computation of vibrational level lifetimes, incorporating both tunneling through the potential barrier and transitions induced by spin-orbit coupling. \\

In short, in the DVR method the vibrational wavefunction is represented on a
discrete grid of points (nodes) in a given interval, \textit{i.e.} effectively
expanded in a truncated orthonormal Hilbert space basis made of DVR-functions.
Each DVR-function possesses the distinct characteristic of being localized on a
single node of the grid and zero at all other nodes. We use here the Lobatto
shape functions introduced by Manolopoulos and Wyatt
\cite{manolopoulos1988quantum}. Finite-Element DVR partitions the space into
contiguous subdomains (elements) while ensuring the continuity of the
vibrational wavefunction on the entire domain. In this manner, the wavefunction
is solved numerically in each element independently by taking advantage of
features of both DVR - diagonal representation of the potential operator - and
FEM - the Hamiltonian matrix is block diagonal and thus broken into smaller pieces \cite{rescigno2000numerical}.
However, the FEM-DVR method requires the vanishing of the wavefunctions at
first and last nodes. In order to address this apparent issue when dealing with
resonant wavefunctions and scattering functions, the method is implemented with
exterior complex scaling~\cite{simon79,nicolaides78}. It involves rotating the real coordinate $R$ into the complex plane by an angle $\theta$ beyond a certain value $R_0$, which in turn exponentially damps the wavefunctions of resonances as $R$ increases. The resulting eigenvalues are complex and of the form:
\begin{equation}
    \varepsilon^{\text{CS}} \:=\: \text{E} - \frac{i}{2}\Gamma,
\end{equation}
where E is the energy of the vibrational level and $\Gamma$ its width.\\

The spatial grid is partitioned as follows: a first element of 400 nodes runs
from 1.7 to 6 $a_0$ and a second element of 200 nodes runs from 6 to 15 $a_0$.
Scaling of the coordinate by the complex-rotation factor starts at the adjacent
node between the two elements, \textit{i.e.} at $R_0=6 \: a_0$, and $\theta$ is initially set at 10$^\circ$. Numerical
convergence was verified by systematically increasing the number of nodes and by systematically 
varying $R_0$ and $\theta$. The results for the lowest vibrational levels of the three
$\Omega$ components of X$^3\Pi$ and of A$^3\Sigma^+$ are presented in Table
\ref{table_vib}. Lifetimes are omitted when the corresponding state width is
too small to fall within the range of accuracy of the method. While predictions
indicate lifetimes between 10$^{-5}$ and 10$^{-3}$ seconds for
X$^3\Pi_{|\Omega|=0,1,2}(v=0,1)$, the reliability of these predictions remains
uncertain as they strongly depend on the grid parameters we noticed. This was also pointed out by others~\cite{mrugala2008computational}.
The angular momentum quantum
numbers $J$ and $N$ are set to the lowest permissible value for each electronic
state ($J \geq\left|\Omega\right|$\cite{herzberg1950electronic}). A strong
dependence of the lifetime on the vibrational quantum number $v$ is observed.
In contrast, the lifetimes of levels with the same $v$ value across different
$\Omega$ components remain  close.

\begin{table}
\caption{\label{table_vib}Energy and lifetime of lowest vibrational levels of \ce{CO^{2+}}. The angular momentum quantum number is set to the lowest permissible value for each electronic state. The energy reference is set to the minimum of the well of X$^3\Pi_{|\Omega|=1}$.}
\begin{ruledtabular}
\begin{tabular}{cccccc}
 & & & $v$ & $E$ (cm$^{-1}$) & $\tau$ (s) \\
\hline
X$^3\Pi$ & $\Omega$=0 & $J$=0 & 0 & 778.0 & $-$\\
         &   && 1 & 2186.6 & $-$ \\
         &   && 2 & 3514.5 & 2.4 $\times10^{-8}$ \\
         &   && 3 & 4806.5 & 1.7 $\times10^{-10}$ \\
         & $\Omega$=1 & $J$=1 & 0 & 716.6 & $-$ \\
         &   && 1 & 2109.7 & $-$ \\
         &   && 2 & 3452.7 & 2.9 $\times10^{-8}$ \\
         &   && 3 & 4741.7 & 1.9 $\times10^{-10}$ \\
         & $\Omega$=2 & $J$=2 & 0 & 659.2 & $-$ \\
         &   && 1 & 2050.3 & $-$ \\
         &   && 2 & 3395.3 & 3.5 $\times10^{-8}$ \\
         &   && 3 & 4687.2 & 2.1 $\times10^{-10}$ \\
\hline
A$^3\Sigma^+$ &   & $N=0$ & 0 & 21132.1 & 5.9 $\times10^{-9}$ \\
              &   && 1 & 23135.1 & 8.1 $\times10^{-10}$ \\
              &   && 2 & 25085.1 & 4.4 $\times10^{-10}$ \\
              &   && 3 & 26979.3 & 2.4 $\times10^{-10}$ \\
\end{tabular}
\end{ruledtabular}
\end{table}

\begin{figure}
 \includegraphics{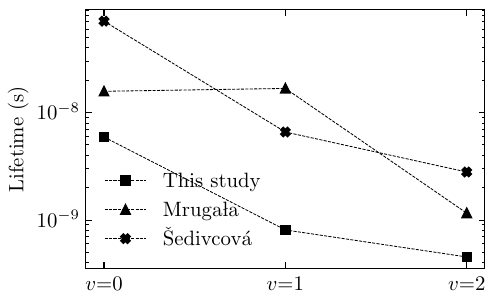}%
 \caption{\label{lft_comp} Lifetime of lowest vibrational levels of A$^3\Sigma^+$ together with the values obtained by Mrugała \cite{mrugala2008computational} and by Šedivcová \textit{et al.} \cite{vsedivcova2007radiative}.}
 \end{figure}

The lifetime of the lowest vibrational levels of A$^3\Sigma^+$ are diplayed on
Fig. \ref{lft_comp}, together with the values reported by Mrugała
\cite{mrugala2008computational} and by Šedivcová \textit{et al.}
\cite{vsedivcova2007radiative}. The values calculated in the present study lie
in the same order of magnitude to those reported in both previous studies.
However, the systematic decrease in lifetime with increasing $v$ was not
reproduced in \cite{mrugala2008computational}. In order to provide an explanation for the decreasing
trend, we computed the tunneling lifetimes, therefore neglecting SOMEs, in order to assess the
effect of spin-orbit coupling on the lifetime relative to tunneling. The LEVEL
program \cite{le2017level} was used as it is renowned for its numerical stability even
for states with very small widths and thus long lifetimes. For all
four vibrational levels $v$=0 to $v$=3 of A$^3\Sigma^+$, the computed lifetimes
were exceedingly long ($>10^{50} $s). This confirms that the lifetimes of such
low-lying vibrational states are predominantly governed by spin-orbit coupling
rather than tunneling. The systematic decrease in lifetime with increasing $v$
can be attributed to the increasing overlap between the vibrational
wavefunctions and those of adjacent electronic states. Indeed, according to Eq.
\ref{schrodinger_total}, the off-diagonal elements of the total Hamiltonian,
which govern the coupling between states, are given by $\langle\chi^{(i)}_k(R) |
\text{A}^{\text{SO}}_{ij}(R) | \chi^{(j)}_l(R)\rangle$. This overlap is minimal between the
wavefunction of A$^3\Sigma^+(v=0)$ and the scattering wavefunctions of the
repulsive state 1$^3\Sigma^-$, as A$^3\Sigma^+(v=0)$ lies below the crossing
point between the two PECs. However, as $v$ increases, this overlap becomes more significant, along with the overlap with the (scattering) wavefunctions of X$^3\Pi$ and b$^1\Pi$, for which the  SO coupling with A$^3\Sigma^+$ is relatively strong, as highlighted in Fig.~\ref{aso}.
\\
The lifetime values that we obtained, presented in Fig. \ref{lft_comp}, are slightly lower than those previously reported. Šedivcová \textit{et al.} employed their own set of PECs and $\text{A}^{\text{SO}}_{ij}(R)$ functions \cite{vsedivcova2006computed}, whereas Mrugała utilized the PECs from Eland \textit{et al.} \cite{eland2004photo} and the $\text{A}^{\text{SO}}_{ij}(R)$ functions from Šedivcová \textit{et al.} \cite{vsedivcova2006computed} for the computation of the lifetime of the lowest vibrational levels of A$^3\Sigma^+$. As previously discussed, the coupling between vibronic states depends on the overlap of vibrational wavefunctions, making lifetime calculations highly sensitive to the accuracy of the underlying PECs, particularly the location of their crossing. As the different sets of PECs slightly differ from each other, as shown in Table \ref{spec_cst}, the inter-state vibrational overlap will differ too and therefore necessarily affects the computed lifetimes.

\subsection{Rotational levels}
The X$^3\Pi_{|\Omega|=0,1,2}$ ground electronic state is well described by Hund's angular-momentum-coupling case (a), whereas the A$^3\Sigma^+$ state is best described by Hund's angular-momentum-coupling case (b)~\cite{nikitin1994correlation} because of the absence of a diagonal spin-orbit interaction. The energies of the rotational levels of the X state are given by~\cite{lefebvre2004spectra}
\begin{equation}
    E^\text{X}(vJ) = T_v^\text{X} + B^\text{X}_v \left[ J (J+1) - \Omega^2 \right] + D^\text{X}_v \left[ J (J+1) - \Omega^2 \right]^2 ,
\end{equation}
with $J$ the quantum number associated with the total angular momentum without nuclear spin and $B_v^X$, $D_v^X$ the rotational constants of the $v$-th vibrational level. The levels of the A state are given by
\begin{equation}
    E^{\text{A}}(vN) = T_v^\text{A} + B^{A}_v \left[ N (N+1) - \Lambda^2 \right], 
\end{equation}
with $N$ the total angular momentum without electron spin. The rotational constants were computed with the LEVEL program \cite{le2017level} and are reported in Table~\ref{tab:vib_constants}.
 
\begin{table}
\caption{Rotational constants $B_v$ and $D_v$ (cm$^{-1}$) for the lowest vibrational levels of the X$^3\Pi$ and A$^3\Sigma^+$ electronic states of \ce{CO^{2+}}.}
\label{tab:vib_constants}
\begin{ruledtabular}
\begin{tabular}{p{0.1cm}cccc}
       & $v$ & $|\Omega|$ & $B_v$ & $D_v$ \\ 
      \cline{2-5}
      X$^3\Pi$ & 0 & 0,1,2 & 1.5675 & 7.7418$\times 10^{-6}$ \\[0.7ex]
      A$^3\Sigma^+$ & 0 & 1 & 1.8992 & 6.7173$\times 10^{-6}$ \\
       & 1 & 1 & 1.8747 & 6.9268$\times 10^{-6}$  \\
       & 2 & 1 & 1.8488  & 7.2016$\times 10^{-6}$  \\
      \end{tabular}
\end{ruledtabular}
\end{table}

We assume that the rotational states of $e$ and $f$ symmetries are degenerate
for both the X state and the A state based on the extensive study of
Ref.~\cite{mrugala2008computational}, which showed that such spin-orbit-induced
splittings are below $0.5$~cm$^{-1}$ for the states considered in the present
paper.



\begin{figure}
    \centering
    \textbf{(a)}
    \includegraphics[width=\columnwidth]{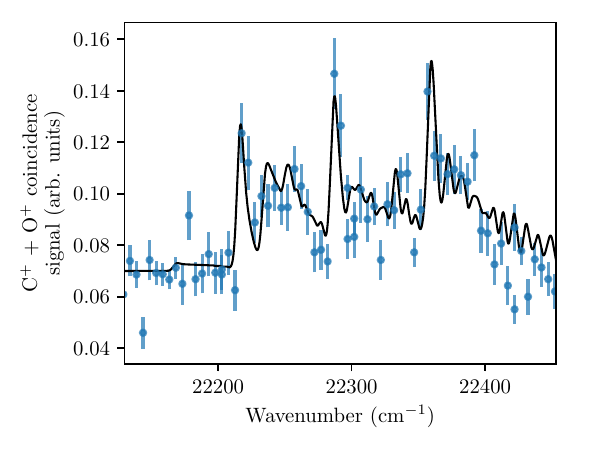}

    \textbf{(b)}
    \includegraphics[width=\columnwidth]{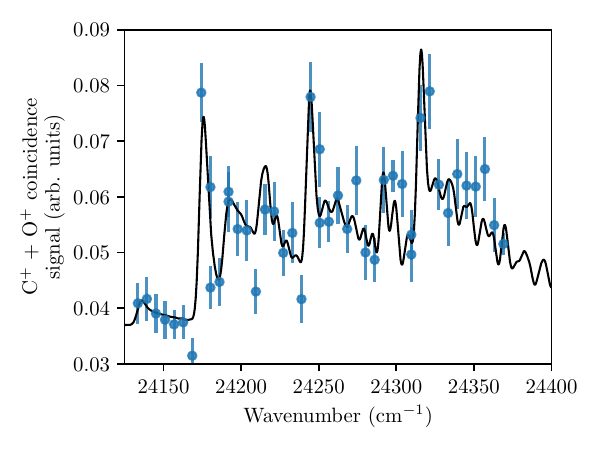}

    \caption{Experimental (blue full circles) and simulated (black full line) of (a) the $\text{A}(v'=1) - X_{\Omega}(v=0)$ band, and (b) the $\text{A}(v'=2) - X_{\Omega}(v=0)$ band. The wavenumber scale of the experimental spectrum is already corrected for the Doppler effect (19~cm$^{-1}$).}
    \label{fig:v1spec}
\end{figure}

\begin{figure}[ht]
    \centering
    \includegraphics[width=\columnwidth]{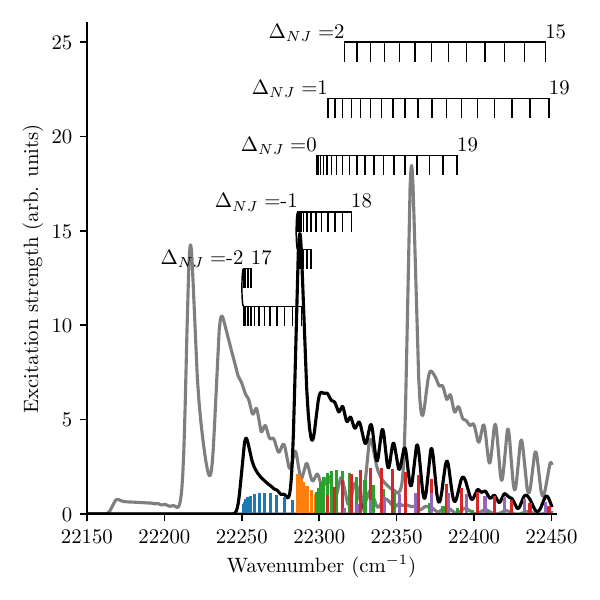}
    \caption{Simulated spectra of each fine-structure component of the $\text{A}(v'=1) - X_{\Omega}(v=0)$ band. Stick spectra and assignment bars are shown for $|\Omega|=1$.}
    \label{fig:assignmentplot}
\end{figure}

\begin{table}
\caption{Band origins of the X$^3\Pi_{|\Omega|}(v=0)$ $\leftarrow$ A $^3\Sigma^+(v')$ transitions ($v'=0-2$ and $|\Omega|=0-2$) in cm$^{-1}$. Numbers in parenthesis give the uncertainty in units of the last digit. The spin-orbit splitting of the X state is unresolved in experimental data from Refs.~\cite{lundqvist1995novel,dawber1994threshold}.}\label{tab:band_origins}
\begin{ruledtabular}
\begin{tabular}{p{0.1cm}cccccc}
      & & \multicolumn{3}{c}{Experiment}  & \multicolumn{2}{c}{Theory}\\ \cline{3-5}\cline{6-7}
      $v'$ & $|\Omega|$ & Present & \cite{lundqvist1995novel} & \cite{dawber1994threshold} & Present & \cite{mrugala2008computational} \\\cline{3-5}\cline{6-7}
      0 & 2 & 20346(3) & \multirow{3}{*}{20245(80)\footnote{The levels X$(v\geq 2)$ were observed in Ref.~\citenum{lundqvist1995novel}. To determine the band origin from X($v=0$) we used the energy difference X($0$)-X($2$) from Ref.~\citenum{dawber1994threshold}.}} & \multirow{3}{*}{20422(80)\footnote{In Ref.~\citenum{dawber1994threshold} this band origin is attributed to $v'=1$. We follow later suggestion by some of the same authors, and by other authors, to reassign it to $v'=0$~\cite{lundqvist1995novel,eland03}}} & 20476.1 & 20668.0 \\
        & 1 & 20281(3) &  &  & 20417.1 & 20603.1 \\
        & 0 & 20218(3) &  &  & 20354.1 & 20540.0 \\[0.15cm]
      1 & 2 & 22361(3) & \multirow{3}{*}{22342(80)} & \multirow{3}{*}{22414(80)} & 22479.1 & 22672.0\\
       & 1 & 22296(3) &  &  & 22420.1 & 22607.1\\
       & 0 & 22233(3) &  &  & 22357.1 & 22544.0\\[0.15cm]
      2 & 2 & 24320(3) & \multirow{3}{*}{24277(80)} & \multirow{3}{*}{24358(80)} & 24429.1 & 24623.1 \\
       & 1 & 24255(3) &  &  & 24370.1 & 24558.1 \\
       & 0 & 24192(3) &  &  & 24307.1 & 24495.1 \\
      \end{tabular}
\end{ruledtabular}
\end{table}

\begin{table}
\caption{Term values of the vibrational levels of the A $^3\Sigma^+(v')$ state in cm$^{-1}$. Numbers in parenthesis give the uncertainty in units of the last digit.}\label{tab:term_values}
\begin{ruledtabular}
\begin{tabular}{p{0.1cm}ccccc}
      & \multicolumn{3}{c}{Experiment}  & \multicolumn{2}{c}{Theory}\\ \cline{2-4}\cline{5-6}
      $v'$ & Present & \cite{lundqvist1995novel} & \cite{dawber1994threshold} & Present & \cite{mrugala2008computational} \\\cline{2-4}\cline{5-6}
      1 & 2015(3) & 2097(80) & 1992(80) & 2003.0 & 2004.0\\
      2 & 3974(3) & 4033(80) & 3936(80) & 3953.0 & 3955.1 \\
      \end{tabular}
\end{ruledtabular}
\end{table}

\section{Results}\label{sec:results}

The experimental spectrum in the region of the $X(v=0)-A(v=1)$ and
$X(v=0)-A(v=2)$ band are shown in Fig.~\ref{fig:v1spec}. Sharp lines attributed
to $X(v=0)-A(v=1,2)$ rovibronic transitions are sumperimposed to a continuous
background attributed to direct photodissociation from $X(v=0)$ into the
continuum above the outer barrier of the $X$-state potential-energy curve (see
Ref.~\cite{vranckx15} for a calculation of the cross section). The kinetic
energy of the fragments recorded at photon energies when only the background is
present is also compatible with direct dissociation via the $X$ state.

A simulation of the $X(v=0)-A(v=1)$ rovibronic spectrum is shown as the black
full line in Fig.~\ref{fig:v1spec}. The spectrum is vertically offset by a
constant to account for the direct photodissociation background. The simulation
is for a rotational temperature of $T_\text{rot}=250$~K, a population of the
initial spin-orbit components of the $X$ state according to their degeneracies,
and rotational constants for the lower- and upper rotational levels given in
Table~\ref{tab:vib_constants}. Rotational line strengths are those for Hund's coupling case (a)
to Hund's coupling case (b) transitions~\cite{zare88}. The calculated spectrum is
convoluted by a gaussian (FWHM~5~cm$^{-1}$) to reproduce the effect of the
finite experimental resolution.

The value of the rotational temperature is suprisingly low because the
duoplasmatron plasma source used to produce CO$^{2+}$ operates at temperatures
of, typically, several thousand kelvins. The electronic
temperature is, as expected, large as all three spin-orbit components of
the X state are significantly populated. A production mechanism in the plasma
discharge yielding rotationally-cold CO$^{2+}$ is unlikely, in particular since
ions prepared with discharges of both pure CO and CO$_2$ gases yielded similar
temperatures. The dependency of the predissociation lifetime of X($v=0$) levels
on the rotational quantum number is theoretically predicted to be
weak~\cite{mrugala2008computational}, which would discard any significant
predissociation of rotationally-hot molecules during the transport of the ions
from the source to the region of interaction with the laser. The origin of the
low apparent rotational temperature is therefore unknown.

The agreement between the experimental and simulated spectra is good and makes it possible to assign the experimental spectrum and extract molecular constants from
it. The assignment is detailed in Fig.~\ref{fig:assignmentplot} for the $|\Omega|=1$ spin-orbit component of the X state. Selection rules for electric dipole rovibronic transitions between the X and A states give rise to five rotational branches corresponding to $\Delta_{NJ} =
N-J'=-2, -1, 0, 1$ and $2$, respectively. The sharpest feature of the $|\Omega|=1$ component is associated
to the bandhead of the $\Delta_{NJ}=-1$ branch, and the same assignment holds for the $|\Omega|=2$ and $0$ components as well. The broad structure to the high-energy side of the sharp line is attributed to
the unresolved rotational structure of the $\Delta_{NJ}>0$ bands.

The presence of one sharp rotational bandhead per fine-structure component of
the X state makes it possible to accurately determine the band origins of the
A($v'$) - X$_\Omega(0)$ transitions despite the rotational structure being not
fully resolved. For a given vibrational band, the spectra for $|\Omega|=0, 1$
and 2 were calculated independently and their band origins fitted to the
experimental spectra. The ordering of the components in energy, with
$|\Omega|=2$ being at the highest wavenumber and $\Omega=0$ at the lowest, was
chosen based on the present \textit{ab initio} calculations, which fall in agreement
with previous works~(see
Refs.~\citenum{vsedivcova2006computed,mrugala2008computational} for recent
examples). Rotationally resolved measurements would allow to confirm this, in
the future, as the $|\Omega|=2$ component should not exhibit any $N=0
\leftarrow J=0$ line. The uncertainty on the band origins is estimated to be
$3$~cm$^{-1}$ and results from the linewidth of the spectral features, which is
mainly due to the spectral width of the laser ($\sim 3$~cm$^{-1}$), the
absolute accuracy of the wavenumber calibration ($1$~cm$^{-1}$), the uncertainty on the
correction of the Doppler shift ($\sim 0.4$~cm$^{-1}$), and the influence of
the value of rotational constants on the bandhead position (2~cm$^{-1}$). To
determine the latter, the rotational constants were taken from the \textit{ab initio}
calculation and varied by $0.1$~cm$^{-1}$ while monitoring the wavenumber of
the maximum of the bandhead. The band origins determined from the experimental
spectra are listed in Table~\ref{tab:band_origins} along with those determined
from the present \textit{ab initio} calculations and compared against previous
experimental~\cite{lundqvist1995novel,dawber1994threshold} and theoretical
values~\cite{mrugala2008computational}. The assignment of a band to a given $v$
is based on the theoretical calculations and previous assignments of
experimental spectra~\cite{lundqvist1995novel,eland03}.

Importantly, the present data provide an experimental determination of the
spin-orbit splitting of the ground electronic state of CO$^{2+}$ as a result of
the 25-fold improvement in resolution compared to previous experimental
spectra~\cite{lundqvist1995novel,dawber1994threshold}. The values of
splittings, near $60$~cm$^{-1}$, are in excellent agreement with present and
past theoretical calculations~\cite{vsedivcova2006computed,mrugala2008computational,cossart99}. The present \textit{ab initio} calculations is also in excellent agreement with experimental term values of the A state (Table~\ref{tab:term_values}), and predicts most accurately the band origins (Table~\ref{tab:band_origins}). This makes us confident that our calculations provide accurate PECs and are able to accurately reproduce the spin-orbit interaction in
CO$^{2+}$.

The experimental band origins also fall in good agreement with
lower-resolution threshold photoelectrons coincidence (TPEsCO) spectra recorded
with synchrotron radiation~\cite{dawber1994threshold} and Doppler-free
kinetic energy release (DFKER) spectra~\cite{lundqvist1995novel}. Neither was
recorded at a resolution sufficient to resolve the spin-orbit splitting of the
X state. The data in Table~\ref{tab:band_origins} further reveal that no spectral lines associated to A-X($v=0$)
transitions can be observed below $\sim 20200$~cm$^{-1}$, and thus confirm the
unsuccessful extensive searches for such transitions conducted at lower wavenumbers by
several groups~\cite{eland03,cox2003high}.

Our results also shed some light
on the tentative identification of CO$^{2+}$ emission lines made by Cossart and
Robbe~\cite{cossart1999comment} and challenged by later theoretical
work~\cite{vsedivcova2007radiative, mrugala2008computational}. Two lines
observed in the Fourier-transform spectrum of a CO discharge at a wavenumber of
$20310.19$~cm$^{-1}$ and $20 368.84$~cm$^{-1}$, respectively, were tentatively
assigned in Ref.~\citenum{cossart1999comment} to the A$(v'=0, N'=1)$ -
X$_{\Omega=0}(v=0, J=1)$ and A $(v'=0, N'=1)$ - X$_{|\Omega|=1}(v=0, J=1)$
transitions, respectively. No other CO$^{2+}$ lines were observed. The
existence of emission lines implies that the corresponding A-state
rovibrational levels are sufficiently long-lived for spontaneous emission to
occur, \textit{i.e.}, that their predissociation lifetimes are longer than the radiative
ones. Such a possibility was theoretically investigated
in~\cite{vsedivcova2007radiative}, however only a single rotational level of
A($v'=0$) was found to have a predissociation lifetime ($11.4$~$\mu$s)
comparable to its radiative lifetime ($7.35$~$\mu$s). A later
study~\cite{mrugala2008computational} found no sufficiently long-lived
rotational levels for emission spectra to be observed, attributing the
long-lived level of~\cite{vsedivcova2007radiative} to a numerical artifact. Our
experimental results do not support the tentative assignment
of~\cite{cossart1999comment} as the lines they analyzed lie more than
90~cm$^{-1}$ above those reported here (Table~\ref{tab:band_origins}).

Finally, the present experimental results place an upper bound of $\sim 110$~ns on the lifetime of the A($v = 1-2$) excited
states. The detection scheme indeed relies on the spontaneous dissociation of the dication after their photoexcitation and
before they reach the detectors, such that the C$^+$ and O$^+$
pair can be detected in coincidence. The width of the measured KER distribution of these fragments remains constant across the spectra shown above, which implies that the fragments are produced right in the region of interaction with the laser and not downstream. If this were to be the case the width would be larger when the fragments are produced via excitation to A($v = 1-2$) resonances compared to when they are produced off resonance via direct photodissociation only. The 3-cm interaction region corresponds to a flight time of $\sim 110$~ns, from which we can deduce that the lifetime of the excited states cannot significantly exceed this value, in agreement with the theoretical results
(Fig. \ref{lft_comp}). Smaller statistics in the case the weaker A($v = 0$) - X($v'=0$) transition do not allow to give a reliable upper bound on the lifetime of the ground vibrational level of the A state.

\section{Conclusion}

We have measured the spectra of the X $^3\Pi_{|\Omega|}(v=0)$ - A
$^3\Sigma^+(v')$ rovibronic transitions ($v'=0-2$) of the CO$^{2+}$ dication.
The 25-fold improvement of the spectral resolution compared to earlier studies
allowed to determine band origins and terms values with a $\sim$~1cm$^{-1}$
accuracy. The spin-orbit structure of the X $^3\Pi_{|\Omega|}(v=0)$ ground
vibronic state is fully resolved and the measured splittings ($\sim
64$~cm$^{-1}$) validate \textit{ab initio} results. The theoretical
calculations reported above fall in good agreement with our experimental investigation, and together provide an accurate description of CO$^{2+}$ that span both structure and nonadiabatic predissociation dynamics of this benchmark
species.

The present experimental approach is versatile and makes it possible to both record spectra at higher resolution, the limit in the present study being the spectral width of the laser, and to study other dications. The lack of high-resolution spectroscopic data for even some of the simplest dications~\cite{thissen2011doubly,cox2003high} is an exciting perspective for future work. Current uncertainties regarding the lifetimes of some vibronic levels of CO$^{2+}$~\cite{mrugala2008computational,cheng25} could also benefit from the recent development of cryogenic storage rings~\cite{vonhahn16,schmidt13}.

The present experimental approach can further be extended to polyatomic molecules, in which case cooling of the rotational degrees of freedom would be important to limit spectral congestion. This represents an  interesting perspective as high-resolution spectra of metal-containing molecules naturally occurring in the doubly-charged state in crystals and liquids would enable to benchmark theoretical calculations and models.

\begin{acknowledgments}
This work was supported by the Fonds de la Recherche Scientifique—FNRS under  IISN Grant No. 4.4504.10, under FRIA Grant No. 40029551 (X. Huet) and under Postdoctoral Researcher Grant No. 40032227 (A. Aerts). Computations were notably performed on the computers of the Consortium des Équipements de Calcul Intensif (C\'ECI).
\end{acknowledgments}

\section*{Data Availability Statement}
The data that support the findings of this study are openly available in Zenodo at \url{http://doi.org/10.5281/zenodo.17406995}, reference number \cite{zenodo_rep}.
\bibliography{ref,paper_MAGE}

\end{document}